\documentclass[12pt]{article}
\usepackage{epsfig,citesort}

\begin{document}

\title{Bell's inequality tests with meson--antimeson pairs}

\author{A.~Bramon$^1$, R.~Escribano$^{1,2}$ and G.~Garbarino$^3$}

\date{\small \it $^1$Grup de F{\'\i}sica Te\`orica, Universitat
Aut\`onoma de Barcelona, E--08193 Bellaterra, Barcelona, Spain \\
$^2$IFAE, Universitat Aut\`onoma de Barcelona, E-08193 Bellaterra, Barcelona, Spain \\
$^3$Dipartimento di Fisica Teorica, Universit\`a di Torino and INFN, Sezione
di Torino, I--10125 Torino, Italy}        

\maketitle

\begin{abstract}
Recent proposals to test Bell's inequalities with entangled
pairs of pseudoscalar mesons are reviewed. This includes pairs of neutral kaons or
$B$--mesons and offers some hope to close both the locality and the detection
loopholes. Specific difficulties, however, appear thus invalidating most of those
proposals. The best option requires the use of kaon regeneration effects and could 
lead to a successful test if moderate kaon detection efficiencies are achieved.
\end{abstract}

%\pacs{PACS numbers: 03.65.Ud, 14.40.Aq}
% 03.65.-w   Quantum mechanics
% 03.65.Ud   Entanglement and quantum nonlocality (e.g. EPR paradox,
%            Bell's inequalities, GHZ states, etc.)
% 14.40.Aq   Pi, K, and eta mesons

%%%%%%%%%%%%%%%%%%%%%%%%%%%%%%%%%%%%%%%%%%%%%%%%%%%%%%%%%%%%%%%%%%%%%%%%%%%%%%%%%
\section{Introduction}
%%%%%%%%%%%%%%%%%%%%%%%%%%%%%%%%%%%%%%%%%%%%%%%%%%%%%%%%%%%%%%%%%%%%%%%%%%%%%%%%%
 
The correlations shown by the distant parts of certain composite systems
offer one of the most counterintuitive and subtle aspects of quantum mechanics.
This was already evident in 1935, when Einstein, Podolsky and 
Rosen (EPR) \cite{epr}, discussing a gedanken experiment with entangled states, 
arrived at the conclusion that the description of physical reality given by the
quantum  wave function cannot be complete. Bohr, in his famous response
\cite{Bo35}, noted that EPR's criterion of physical
reality contained an ambiguity if applied to quantum phenomena:
an argument using the complementarity point of view led him to conclude that quantum
mechanics, in the form restricted to human knowledge, ``would appear as a completely rational 
description of the physical phenomena''.

For about 30 years the debate triggered by EPR and Bohr
remained basically a matter of philosophical belief. 
Then, in 1964, Bell \cite{bell} interpreted EPR's argument as the need for the introduction
of additional, unobservable variables aiming to restore {\it completeness},
{\it relativistic causality} (or {\it locality}) and {\it realism} in quantum
theory. He established a theorem which proved that any {\it local
hidden--variable}  (i.e., {\it local realistic}
\cite{Redhead}) theory is incompatible with some  statistical predictions of
quantum mechanics. Since then,  various forms of Bell's inequalities
\cite{wig,Bell2,chsh,clh} have been the tool for  an experimental discrimination
between local realism (LR) and quantum mechanics (QM).

Many experiments have been performed, mainly with entangled photons 
\cite{As82,review1,We98,Ti99,BZ} and ions \cite{Ro01}, in order to confront LR
with QM.  All these tests obtained results 
in good agreement with QM and showed the violation of {\it non--genuine}
Bell's inequalities. Indeed, because of non--idealities of the apparata 
and other technical problems, {\it supplementary assumptions} not implicit in LR were 
needed in the interpretation of the experiments. Consequently, no one of these experiments
has been strictly {\it loophole free} \cite{review1,vaidman,santos}, i.e., able to
test a {\it genuine} Bell's inequality.

It has been proven \cite{clh,review1,pearle} that for any entangled state
one can derive Bell's inequalities without the introduction of
(plausible but not testable) supplementary assumptions concerning undetected events.
For maximally entangled (non--maximally entangled) states, if one assumes
that all detectors have the same overall efficiency $\eta$, these genuine inequalities
are violated by QM if $\eta >0.83$ \cite{GM} ($\eta > 0.67$ \cite{Eb}).
Since such thresholds cannot be presently achieved in photon experiments,
only non--genuine inequalities have been tested experimentally.
They are then violated by QM irrespectively of the detection efficiency values.

Several of these photonic tests violated non--genuine inequalities
by the amount predicted by QM but they could not overcome the {\it detection loophole}.
Indeed, local realistic models exploiting detector inefficiencies
and reproducing the experimental results can be contrived \cite{clh,santos-models}
for these tests. Only the recent experiment with entangled beryllium ions of Ref.~\cite{Ro01},
for which $\eta\simeq 0.97$, did close the detection loophole.
But then the other existing loophole, the {\it locality loophole},
remains open due to the tiny inter--ion separation.
Conversely, an experiment with distant entangled photons \cite{We98} closed this latter loophole.
In this test, 
the measurements on the two photons were carried out under
space--like separation conditions, thus avoiding any exchange of subluminal signals between the
two measurement events, but detection efficiencies were too low to close the detection loophole.
In other words, no experiment closing simultaneously both loopholes has been performed till now.

Extensions to other kinds of entangled systems are thus important.
Over the past ten years or so there has been an increased interest on
the possibility to test LR vs QM in particle physics, i.e., by using entangled neutral kaons
\cite{ghi,eberhard,domenico,uchiyama,selleri,AS,bf,berthies,bn,abn,gigo,dg,gnp,BG,BG1,hies-th,genovese,Bert}
or $B$--mesons \cite{DH,Se00,Go,antiGo,antiGo2}.
This is also a manifestation of the desire to go beyond
the usually considered spin--singlet case 
and to have new entangled systems made of massive particles
with peculiar quantum--mechanical properties.
Entangled $K^0\bar K^0$ ($B^0\bar B^0$) pairs are
produced in the decay of the $\phi$ resonance \cite{daphne}
($\Upsilon(4S)$ resonance \cite{belle}) and in proton--antiproton annihilation
processes at rest \cite{CPLEAR}. For kaons, the strong
nature of hadronic interactions should contribute to close the detection loophole,
since it enhances the efficiencies 
to detect the products of kaon decays and kaon interactions with ordinary matter 
(pions, kaons, nucleons, hyperons,...). Moreover, 
the two kaons produced in $\phi$ decays or $p\bar{p}$ annihilations at rest fly
apart from each other at relativistic velocities and easily fulfill the condition
of space--like separation. Therefore, contrary to the experiment with ion pairs of
Ref.~\cite{Ro01}, the locality loophole could be closed with
kaon pairs by using equipments able to prepare, very rapidly, the
alternative kaon measurement settings.

In this contribution our purpose is to review the Bell's inequalities
proposed to test LR vs QM using entangled pairs of neutral pseudoscalar mesons 
such as $K^0\bar K^0$ and $B^0\bar B^0$. 
These proposals will be discussed on the light of the basic requirements 
necessary to establish genuine Bell's inequalities.

%%%%%%%%%%%%%%%%%%%%%%%%%%%%%%%%%%%%%%%%%%%%%%%%%%%%%%%%%%%%%%%%%%%%%%%%%%%%%%%%%
\section{Neutral meson systems}
%%%%%%%%%%%%%%%%%%%%%%%%%%%%%%%%%%%%%%%%%%%%%%%%%%%%%%%%%%%%%%%%%%%%%%%%%%%%%%%%%
\label{nms}
%%%%%%%%%%%%%%%%%%%%%%%%%%%%%%%%%%%%%%%%%%%%%%%%%%%%%%%%%%%%%
\subsection{Single mesons: time evolution and measurements}
%%%%%%%%%%%%%%%%%%%%%%%%%%%%%%%%%%%%%%%%%%%%%%%%%%%%%%%%%%%%%
\label{sm}
In this section we discuss the time evolution of and the kind of measurements on 
neutral  pseudoscalar mesons. We mainly refer to the most known case of
neutral kaons, but the modifications
which apply to neutral $B$--mesons are stressed as well.
These differences originate from the different values of the meson parameters
and turn out to have important consequences when testing LR vs QM.

Neutral kaons are copiously produced by strangeness--conserving strong
interaction  processes such as $\pi^- p\to \Lambda K^0$ and $p \bar p\to K^-
\pi^+ K^0$ and so they initially appear either as $K^0$'s (strangeness $S=+1$) or
$\bar{K^0}$'s  (strangeness $S=-1$). 
The distinct strong interactions of the $S=+ 1$ and $S=- 1$ kaons on the bound 
nucleons 
of absorber materials project an incoming kaon state into one of these two orthogonal
members of the strangeness basis $\{K^0,\bar K^0\}$, and permit the measurement of $S$ \cite{AS}. 
This strangeness detection is
totally analogous to the projective von Neumann measurements with two--channel
analyzers for polarized photons or Stern--Gerlach set-ups for spin--$1/2$ particles.
Unfortunately, the detection efficiency for such strangeness measurements
is rather limited \cite{CPLEAR}. 
Indeed, it could be close to $1$ only for infinitely dense absorber materials
or for ultrarelativistic kaons, where, by Lorentz contraction, the absorber
is seen by the incoming kaon as extremely dense. In this case, kaon--nucleon strong 
interactions become
much more likely than kaon weak decays. It would be highly desirable to identify very efficient
absorbers. Since this does not seem to be viable at present, one has to play with small
strangeness detection efficiencies, which 
originate serious conceptual difficulties
when discussing Bell--type tests for entangled kaons \cite{BG,genovese}.

The kaon time--evolution and decay in free space is governed by the lifetime basis, $\{K_S,K_L\}$, 
whose states diagonalize the non--Hermitian weak Hamiltonian. 
The proper time propagation of these short-- and long--lived states 
having well--defined masses $m_{S,L}$ is given by:
\begin{equation}
\label{evo}
|K_{S,L}(\tau)\rangle= e^{-i m_{S,L} \tau}
e^{-{1 \over 2}\Gamma_{S,L} \tau} |K_{S,L}\rangle ,
\end{equation}
where $\Gamma_{S,L}\equiv 1/\tau_{S,L}$ are
the kaon decay widths and $\tau_S=(0.8953\pm 0.0005)\times 10^{-10}$ s and 
$\tau_L=(5.18\pm 0.04)\times 10^{-8}$ s \cite{PDG} the corresponding lifetimes.
Being the dynamics of free kaons governed by strangeness non--conserving weak
interactions, $K^0$--$\bar K^0$ mixing and $K_S$--$K_L$ interferences will appear thus
producing the well known $K^0$--$\bar K^0$ oscillations in time. Assuming
$CPT$ invariance,  the relationship between strong and weak interaction eigenstates is
provided by \cite{kabir}:
\begin{eqnarray}
\label{strong-weak}
|K_S \rangle= 
\frac{1}{\sqrt{2(1+|\epsilon|^2)}}\left[(1+\epsilon)|K^0\rangle+
(1-\epsilon)|\bar{K}^0\rangle \right] , \\
|K_L \rangle= 
\frac{1}{\sqrt{2(1+|\epsilon|^2)}}\left[(1+\epsilon)|K^0\rangle
-(1-\epsilon)|\bar{K}^0\rangle \right] , \nonumber
\end{eqnarray}
$\epsilon$ being the $CP$--violation parameter in the $K^0$--$\bar K^0$ mixing.
Weak interaction eigenstates are related to the $CP$ eigenstates $|K_1\rangle$ 
($CP=+1$) and $|K_2\rangle$ ($CP=-1$) by:
\begin{eqnarray}
\label{cp}
|K_S\rangle &=& \frac{1}{\sqrt{1+|\epsilon|^2}}\left[|K_1\rangle+\epsilon |K_2\rangle\right] , \\
|K_L\rangle &=& \frac{1}{\sqrt{1+|\epsilon|^2}}\left[|K_2\rangle+\epsilon |K_1\rangle\right] .
\nonumber
\end{eqnarray}

To observe if a kaon is propagating as a $K_S$ or $K_L$ at time $\tau$,
one has to identify at which time it subsequently decays. Kaons which show a decay
between times $\tau$ and $\tau + \Delta \tau$ have to be identified as $K_S$'s, while those
decaying later than $\tau + \Delta \tau$ have to be identified as $K_L$'s. The
probabilities for wrong $K_S$ and $K_L$ identification are then given by
$\exp(- \Gamma_S\, \Delta \tau)$ and $1 - \exp(- \Gamma_L\, \Delta \tau)$, respectively.
With $\Delta \tau = 4.8\, \tau_S$, both $K_S$ and $K_L$ misidentification probabilities
reduce to $\simeq 0.8$\%. 
Note that the $K_S$ and $K_L$ states are not strictly orthogonal to each other,
$\langle K_S|K_L \rangle = 2\, {\cal R}e\, \epsilon / (1+|\epsilon |^2 ) \neq 0$,
thus their identification cannot be exact even in principle. However, $\epsilon$ is so
small [$|\epsilon |\simeq (2.284\pm 0.014)\times 10^{-3}$ \cite{PDG}] 
and the decay probabilities of the two
components so different ($\Gamma_S \simeq 579 \, \Gamma_L$) that the $K_S$ vs $K_L$
identification effectively works in many cases \cite{BG}.
Note also that, contrary to strangeness measurements, lifetime observations 
can be made with quite high efficiencies; by using detectors with 
very large solid angles, one can play with
almost ideal efficiencies ($\eta_\tau \simeq 1$) for the detection of the kaon decay
products.

Apart from this (only approximate) $K_S$ vs $K_L$ identification and the previous
(in principle exact) strangeness measurement, no other quantum--mechanical measurement
with dichotomic outcomes is possible for neutral kaons \cite{BG}. Only these two complementary
observables can be exploited to establish Bell's inequalities. This is in sharp contrast to
the standard spin--singlet case and reduces the possibilities of kaon experiments.

The above methods used to discriminate $K^0$ vs $\bar K^0$ and $K_S$ vs $K_L$  
correspond to {\it active} measurement procedures
since they are performed by exerting the free will of the experimenter. Indeed, at a
chosen time, either one places a slab of matter or allows for free  space propagation.
Contrary to what happens with other two--level quantum systems, such as spin--$1/2$
particles or photons, {\it passive} measurements of strangeness and
lifetime for neutral kaons are also possible \cite{BGHPR} by randomly exploiting the
quantum--mechanical dynamics of kaon decays. 

The strangeness content of neutral kaon states can indeed
be determined by observing their semileptonic decay modes, which obey the
well tested $\Delta S=\Delta Q$ rule. This rule allows the modes 
$K^0 \to \pi^- +l^+ +\nu_l$ and $\bar{K}^0 \to\pi^+ + l^- + \bar\nu_l$
($l=e, \mu$) but forbids decays into the
respective charge conjugated modes. Obviously, the experimenter
cannot induce a kaon to decay semileptonically and not even at a given
time: he or she can only sort at the end of the day all observed
events in proper decay modes and time intervals.
Therefore, this discrimination between $K^0$ and $\bar K^0$ is called
a {\it passive} measurement of strangeness. As in the case of active strangeness
measurements, the detection efficiency for passive strangeness measurements is rather limited
---it is given by the $K_L$ and $K_S$ semileptonic branching
ratios, which are $\simeq 0.66$ and $\simeq 1.1\times 10^{-3}$,
 respectively \cite{PDG}. Again, this  poses serious problems when testing LR vs QM.

By neglecting the small $CP$ violation effects ($\epsilon=0$ and thus  
$\langle K_S |K_L\rangle =0$), one can discriminate between 
$K_S$'s and $K_L$'s by leaving the kaons to propagate in free space and by observing
their distinctive nonleptonic $K_S \to 2\pi$ or $K_L \to 3\pi$
decays. This represents a passive measurement of
lifetime, since the kaon decay modes ---nonleptonic in the
present case, instead of semileptonic as before--- as well as the 
decay times cannot be in any way influenced by the experimenter. 

We therefore have two conceptually different experimental
procedures to measure each one of the two neutral kaon observables.
The active measurement of strangeness is monitored by
strangeness conservation while the corresponding passive
measurement is assured by the $\Delta S=\Delta Q$ rule.
Active and passive lifetime measurements are possible
thanks to the smallness of $\Gamma_L/\Gamma_S$ and $\epsilon$, respectively. 
Note that with the passive measurement method, the mere quantum--mechanical dynamics
of kaon decays decides if the neutral kaon is going to be
measured either in the strangeness or in the lifetime basis. The experimenter
remains totally passive in such measurements, which are thus clearly different
from the usual, active von Neumann projection measurements.

Both active and passive procedures lead to the same probabilities for strangeness and 
lifetime measurements \cite{BGHPR}.
Considering the evolution of a neutral kaon produced at $\tau=0$
as a $K^0$, in both cases one easily obtains the following transition probabilities:
\begin{eqnarray}
\label{k0}
P(K^0(0)\to K^0 (\tau ))
&=& {1 \over 4}(e^{-\Gamma_S \tau}+e^{-\Gamma_L \tau}) 
\left[ 1+\frac{\cos\, (\Delta m \, \tau)}{\cosh\, (\Delta \Gamma\, \tau/2)}\right], \\
\label{k0bar}
P(K^0(0)\to \bar{K}^0 (\tau ))
&=& {1 \over 4}(e^{-\Gamma_S \tau}+e^{-\Gamma_L \tau})
\left[ 1-\frac{\cos\, (\Delta m \, \tau)}{\cosh\, (\Delta \Gamma\, \tau/2)}\right] , \\
P(K^0(0)\to K_L (\tau )) &=& {1 \over 2} e^{-\Gamma_L \tau} , \\
\label{ks}
P(K^0(0)\to K_S (\tau )) &=& {1 \over 2} e^{-\Gamma_S \tau},
\end{eqnarray}
where $\Delta m \equiv m_L-m_S$ and $\Delta \Gamma \equiv \Gamma_L-\Gamma_S$
are determined by strangeness oscillation experiments through
Eqs.~(\ref{k0}) and (\ref{k0bar}). The experimental equivalence of 
active and passive measurement procedures on single kaon states and
the agreement with quantum--mechanical predictions have already been
established \cite{PDG,CPLEARreview,BGH2}.

The existence of the two measurement procedures 
---active and passive--- opens new possibilities for tests of basic 
principles of QM with kaons \cite{BGHPR} 
---such as quantum erasure and quantitative formulations of Bohr's
complementarity--- which have no analog for any other two--level
quantum system considered up to date. Unfortunately, as we will see in detail
in Section \ref{bell}, passive measurements are of no interest when testing Bell's
inequalities with kaons, where only {\it active} measurements must be
considered \cite{BG,BG1}.

Neutral $B$--mesons are easily produced at asymmetric $B$--factories
using high luminosity and asymmetric $e^+e^-$ colliders 
operating at the $J^{PC}=1^{--}$ $\Upsilon(4S)$ resonance \cite{belle}. 
For these mesons, the strangeness eigenstates are replaced by the {\it beauty}
eigenstates $|B^0\rangle$ and $|\bar B^0\rangle$, while the light ($m_L$) and heavy
($m_H$) mass eigenstates are $|B_L\rangle$ and $|B_H\rangle$.
Experimentally, we know that $B_L$ and $B_H$ 
have very similar decay widths: $|\Delta \Gamma_B|/ \Gamma_B<0.18$ at 95\% CL, where
$\Delta \Gamma_B=\Gamma_H-\Gamma_L$ and $\Gamma_B= (\Gamma_L+\Gamma_H)/2\equiv 1/\tau_B$, 
with $\tau_B=(1.536\pm 0.014)\times 10^{-12}$ s \cite{PDG}. 
With these changes, Eqs.~(\ref{evo})--(\ref{ks})
still hold if $\epsilon$ ($\Delta m$) is replaced by $\epsilon_{B^0}$ ($\Delta m_B=m_H-m_L$), 
the $CP$ violation parameter in the $B^0$--$\bar B^0$ mixing.
Contrary to the kaon case, $CP$ violation in the
$B^0$--$\bar B^0$ mixing has not been observed unambiguously, since
$\langle B_L|B_H \rangle = 2\, {\cal R}e\, \epsilon_{B^0} / (1+|\epsilon_{B^0}|^2)
=(1.0\pm 6.2)\times 10^{-3}$ \cite{PDG}.
Experimentally one knows that for kaons and $B$--mesons one has $|\Delta \Gamma
|\simeq 2.1\, \Delta m$ and $|\Delta \Gamma_B|\simeq 0.23\, \Delta m_B$, respectively; 
thus, the number of flavour oscillations that one can observe in Eqs.~(\ref{k0}) and
(\ref{k0bar}) is much larger for $B$--mesons than for $K$--mesons.

Concerning neutral $B$--meson measurements, the main 
difference with respect to the neutral kaon case is that {\it active}
flavour (strangeness or beauty) measurement procedures
are only available for kaons \cite{antiGo,antiGo2}. The $B$--meson beauty
can only be determined through a {\it passive} procedure, 
by observing the meson decay modes. The series of decay products 
$f=  D^*(2010)^- l^+ \nu_l$, $D^- \pi^+,\dots$,
which are forbidden for a $\bar B^0$, necessarily come from a
$B^0$, while the opposite is true for the respective charge conjugated modes
$\bar f=  D^*(2010)^+ l^- \bar \nu_l$, $D^+ \pi^-,\dots$ ($l= e, \mu$).
Passive $B$--meson measurements able to distinguish between $B_L$'s and
$B_H$'s are almost impossible to perform nowadays, especially if operated in an 
experiment aiming to test a Bell's inequality, due to the small value
of $|\Delta \Gamma_B|/\Gamma_B$.
As we discuss in Section~\ref{bell}, these limitations play a decisive role when 
testing LR vs QM with entangled $B$--mesons.

%%%%%%%%%%%%%%%%%%%%%%%%%%%%%%%%%%%%%%%%%%%%%%%%%%%%%%%%%%%%%%%%%%%%%%%
\subsection{Entangled meson pairs}
%%%%%%%%%%%%%%%%%%%%%%%%%%%%%%%%%%%%%%%%%%%%%%%%%%%%%%%%%%%%%%%%%%%%%%
\label{emp}

Let us now consider two--kaon entangled states which are analogous
to the standard and widely used two--photon entangled states
\cite{bn,gigo,Bert,BGH}. From both $\phi$--meson resonance decays
\cite{daphne} or $S$--wave proton--antiproton annihilation
\cite{CPLEAR}, one starts at time $\tau=0$ with the $J^{PC}=1^{--}$ state:
\begin{equation}
\label{entangled} 
|\phi(0)\rangle  =  \frac{1}{\sqrt 2}\left[
|K^0\rangle_l |\bar{K}^0\rangle_r - |\bar{K}^0\rangle_l
|K^0\rangle_r\right]
 =  \frac{1}{\sqrt 2}\frac{1+|\epsilon|^2}{1-\epsilon^2}\left[
|K_L\rangle_l |K_S\rangle_r - |K_S\rangle_l |K_L\rangle_r\right] ,
\end{equation}
where $l$ and $r$ denote the ``left'' and ``right'' directions of
motion of the two separating kaons and $CP$--violating effects enter
the last equality. Note that this state is
antisymmetric and maximally entangled in the two observable bases.

After production, the left  and right moving kaons evolve
according to Eq.~(\ref{evo}) up to times $\tau_l$ and
$\tau_r$, respectively. This leads to the state:
\begin{eqnarray}
\label{time}
|\phi(\tau_l , \tau_r)\rangle &=& \frac{1}{\sqrt{2}}
e^{-(\Gamma_L\, \tau_l +\Gamma_S\, \tau_r)/2}
\left\{|K_L\rangle_l |K_S\rangle_r - e^{i \Delta m (\tau_l -\tau_r)} 
e^{\Delta \Gamma (\tau_l -\tau_r)/2} |K_S\rangle_l |K_L\rangle_r\right\} 
\end{eqnarray}
in the lifetime basis, or:
\begin{eqnarray}
\label{statet} 
|\phi(\tau_l , \tau_r)\rangle &=& \frac{1}{2\sqrt{2}} 
e^{-(\Gamma_L\, \tau_l +\Gamma_S\, \tau_r)/2}
\left\{ \left(1-e^{i \Delta m (\tau_l -\tau_r)} e^{\Delta \Gamma (\tau_l -\tau_r)/2}\right)
\left[|K^0\rangle_l|K^0\rangle_r-|\bar K^0\rangle_l|\bar
K^0\rangle_r \right] \right. \nonumber \\  
&&+ \left. \left(1+e^{i \Delta m (\tau_l -\tau_r)} e^{\Delta \Gamma (\tau_l -\tau_r)/2}\right)
\left[|K^0\rangle_l|\bar K^0\rangle_r-|\bar K^0\rangle_l|K^0\rangle_r\right]\right\} 
\end{eqnarray}
in the strangeness basis, where small $CP$ violation effects have been safely neglected. 

Note the analogy between state (\ref{time}) and the polarization--entangled two--photon 
[idler ($i$) plus signal ($s$)] state used in optical tests of Bell's inequalities:
\begin{equation}
|\Psi \rangle = \frac{1}{\sqrt{2}}
\left\{ |V\rangle_i |H\rangle_s - e^{i \Delta \phi}|H\rangle_i |V\rangle_s \right\} , 
\end{equation}
where $\Delta \phi$ is an adjustable relative phase.
For entangled kaons, the non--vanishing value
of $\Delta m$ plays the same role as $\Delta \phi$ and induces 
$K_S$ and $K_L$ interferences, as seen from Eq.~(\ref{time}), as well as 
strangeness oscillations in time. These oscillations can be used to mimic the
different orientations of polarization analyzers in photonic Bell--tests 
\cite{bn,gigo}. 

The same-- and opposite--strangeness detection probabilities:
\begin{eqnarray}
\label{4P1}
P(K^0,\tau_l; K^0, \tau_r) &=&  P(\bar K^0,\tau_l; \bar K^0,\tau_r) \nonumber \\
&=& {1 \over 8} \left(e^{-(\Gamma_L\, \tau_l+\Gamma_S\, \tau_r)}
+e^{-(\Gamma_S\, \tau_l+\Gamma_L\, \tau_r)}\right) 
\left\{ 1- \frac{\cos\, [\Delta m (\tau_l-\tau_r)]}
{\cosh\, [\Delta \Gamma 
(\tau_l-\tau_r)/2]}\right\} , \\
\label{4P2}
P(K^0,\tau_l; \bar K^0,\tau_r) &=&  P(\bar K^0, \tau_l; K^0,\tau_r) \nonumber \\
&=&{1 \over 8} \left(e^{-(\Gamma_L\, \tau_l+\Gamma_S\, \tau_r)}
+e^{-(\Gamma_S\, \tau_l+\Gamma_L\, \tau_r)}\right)
\left\{ 1+ \frac{\cos\, [\Delta m (\tau_l-\tau_r)]}{\cosh\, [\Delta \Gamma (\tau_l-\tau_r)/2]}\right\} ,
\end{eqnarray}
are obtained for both active and passive joint measurements \cite{BGHPR}. Note that 
for entangled $B$--meson pairs created in 
$\Upsilon(4S)\to B^0\bar B^0$ decays,
the same-- and opposite--beauty detection probabilities simplify into:
\begin{eqnarray}
\label{b-m1}
P(B^0,\tau_l; B^0, \tau_r) =  P(\bar B^0,\tau_l; \bar B^0, \tau_r) =
\frac{1}{4} e^{-(\tau_l +\tau_r)\, \Gamma_B} 
\left\{1- \cos [\Delta m_B \, (\tau_l-\tau_r)]\right\} , &&
\\
\label{b-m2}
P(B^0, \tau_l; \bar B^0, \tau_r) =  P(\bar B^0,\tau_l; B^0, \tau_r) =
\frac{1}{4} e^{-(\tau_l +\tau_r)\, \Gamma_B} 
\left\{1+ \cos [\Delta m_B \, (\tau_l-\tau_r)]\right\} , &&
\end{eqnarray}
due to the smallness of the  $B_L$ and $B_H$ lifetime difference
($\Gamma_L=\Gamma_H=\Gamma_B$). Note also that for $\tau_l=\tau_r$ we have perfect
EPR--correlations in $J^{PC}=1^{--}$ meson--antimeson pairs: 
the same--flavour probabilities (\ref{4P1}) and (\ref{b-m1}) vanish and the 
opposite--flavour probabilities (\ref{4P2}) and (\ref{b-m2}) take the maximal values.

Entanglement in the flavour quantum number has been tested experimentally,
over macroscopic distances, for
kaons at CPLEAR \cite{CPLEAR}, using active strangeness measurements,
and for $B$--mesons at Belle \cite{Go}, using passive measurements of beauty.
The non--separability of the meson--antimeson $J^{PC}=1^{--}$ state could be also observed
at the Da${\Phi}$ne $\phi$--factory \cite{daphne},
using passive \cite{kloe-int} and (with some modification of the set--up) active 
strangeness measurements.

%%%%%%%%%%%%%%%%%%%%%%%%%%%%%%%%%%%%%%%%%%%%%%%%%%%%%%%%%%%%%%%%%%%%%%%%%%%%%%%%%
\section{Bell's inequality tests with meson--antimeson pairs}
%%%%%%%%%%%%%%%%%%%%%%%%%%%%%%%%%%%%%%%%%%%%%%%%%%%%%%%%%%%%%%%%%%%%%%%%%%%%%%%%%
\label{bell}
%%%%%%%%%%%%%%%%%%%%%%%%%%%%%%%%%%%%%%%%%%%%%%%%%%%%%%%%%%%%%%%%%%
\subsection{Requirements to establish a genuine Bell's inequality}
%%%%%%%%%%%%%%%%%%%%%%%%%%%%%%%%%%%%%%%%%%%%%%%%%%%%%%%%%%%%%%%%%%
\label{req}
The requirements for deriving a Bell's inequality from LR can be summarized as follows:
\begin{itemize}
\item[(1)] A non--factorizable or entangled state must be used;
\item[(2)] Alternative (mutually exclusive) measurements 
      corresponding to two non--commuting observables must be chosen at will
      both on the left and on the right side;
\item[(3)] To each single measurement corresponds dichotomic outcomes
(or  trichotomic if the possibility of undetected events is considered as a third
outcome);
\item[(4)] Measurement events must be space--like separated.
\end{itemize}

The first requirement poses no problem. As previously stated, 
entanglement has been confirmed experimentally for meson--antimeson pairs.
It is then important to explore the possibility
to derive genuine Bell's inequalities for such systems.

Difficulties appear with requirement number (2). Indeed, 
among the differences between the singlet--spin state of entangled photons and the
$K^0\bar K^0$ entangled state previously considered, 
the most important one is that while for photons one can measure
the linear polarization along {\it any} space direction chosen at will,
measurements on neutral kaons are only of {\it two} kinds: one can 
chose to measure either strangeness or lifetime. 
This reduces considerably the possibilities of Bell--tests with neutral kaons. 
For entangled $B^0\bar B^0$ pairs the situation is even 
more unfortunate: indeed, the lack of
active measurement procedures for $B$--mesons makes impossible the derivation of genuine Bell's 
inequalities \cite{antiGo2}.

Also, in order to establish the feasibility of a real test, one has to derive 
the detection efficiencies necessary for a meaningful quantum--mechanical violation 
of the considered Bell's inequality. In addition, decay events are known to further
complicate the issue. With all this in mind and in
the light of the basic requirements (1)--(4), we proceed now to analyse various
proposals of Bell--tests with entangled meson--antimeson pairs.

%%%%%%%%%%%%%%%%%%%%%%%%%%%%%%%%%
\subsection{Proposals with passive measurements}
%%%%%%%%%%%%%%%%%%%%%%%%%%%%%%%%% 
A recent paper \cite{Go} claims that a violation of a Bell's inequality 
has been observed for the first time in particle physics using the
particle--antiparticle correlations in semileptonic $B$--meson decays.
Other authors \cite{bf} proposed an analogous test with neutral kaons.
%and concluded that it was ``the first experimental  evidence of a
%violation of Bell's locality  occurring in a subnuclear system''. 
In the following we show that, since $B$-- or $K$--decays serve to identify 
flavour passively, the inequalities considered in Refs.~\cite{bf,Go} 
cannot be considered genuine Bell's inequalities. 

To exemplify, let us consider in some detail the recent Belle test \cite{Go}, where
an entangled $B$--meson state analogous to that of Eqs.~(\ref{time}) 
and (\ref{statet}) was employed. The experiment measured the joint probabilities 
of Eqs.~(\ref{b-m1}) and (\ref{b-m2}). The flavour
of each member of the pair was identified by observing its semileptonic
decay. The decay channel $f=  D^*(2010)^- l^+ \nu_l$,
which is forbidden for a $\bar B^0$, unambiguously comes from a
$B^0$, while the opposite is true for the respective charge conjugated mode
$\bar f=  D^*(2010)^+ l^- \bar \nu_l$ ($l= e, \mu$).
The corresponding partial decay widths satisfy
$\Gamma_{B^0 \to f} = \Gamma_{\bar B^0 \to \bar f}$ \cite{PDG}.
Experimentally, one counts the number of joint $B$--meson decay events
into the distinct decay modes $f_{l,r}$ and
in the appropriate time intervals $[\tau_{l,r}, \tau_{l,r} + d\tau_{l,r}]$; then the
joint decay probabilities ${\cal P}(f_l, \tau_l ; f_r, \tau_r)$ are obtained after
dividing these numbers by the total number of initial $B^0\bar B^0$ pairs.
Finally, the corresponding joint decay rates $\Gamma(f_l, \tau_l ; f_r, \tau_r)$ are derived as:
\begin{eqnarray}
\label{gamma}
\Gamma(f_l, \tau_l ; f_r, \tau_r)\equiv
\frac{d^2 {\cal P}(f_l, \tau_l ; f_r, \tau_r)}{d\tau_l\; d\tau_r} = P(B_l, \tau_l ; B_r, \tau_r)\;
\Gamma_{B_l \to f_l}\; \Gamma_{B_r \to f_r} ,
\end{eqnarray}
from which the joint probabilities $P(B_l, \tau_l ; B_r, \tau_r)$ of Eqs.~(\ref{b-m1}) and (\ref{b-m2})
immediately follow. The data of Ref.~\cite{Go} 
are found to be in good agreement with the quantum--mechanical predictions
in Eqs.~(\ref{b-m1}) and (\ref{b-m2}). This is a convincing proof of the entanglement between the 
two members of each $B$--meson pair, but is it a meaningful test confronting LR vs QM?
 
In our view and because of the lack
of active measurements, the Clauser, Horne, Shimony and Holt (CHSH) 
\cite{chsh} inequality tested in Ref.~\cite{Go} is not  a genuine Bell's inequality. 
The conventional and most convincing procedure to demonstrate this consists in
constructing a local model of hidden variables which agrees with the quantum--mechanical
predictions and thus with the experimental data of Ref.~\cite{Go}. In the present
case, this is easily achieved \cite{antiGo2} by simply adapting an original argument
introduced by Kasday \cite{Kasday} in another context. Each $B^0
\bar B^0$ pair is assumed to be produced at $\tau=0$ with a set of hidden variables
$\{\tau_l, f_l, \tau_r, f_r\}$ deterministically specifying {\it ab initio}
the future decay times and decay modes of its two members.
Different $B$--meson pairs are then supposed to be produced with a probability
distribution coinciding precisely with the joint decay probability
${\cal P}(f_l, \tau_l ;f_r, \tau_r)$ entering Eq.~(\ref{gamma}).
Note that the conventional normalization in the hidden variable space,
$\int d\lambda\, \rho (\lambda )=1$, is now similarly given by
$\Sigma_{f_l,f_r} \int d\tau_l \int d\tau_r\, \Gamma(f_l, \tau_l ; f_r, \tau_r) = 1$,
where the time integrals extend from 0 to $\infty$ and the sum to all $B^0$ and $\bar B^0$
decay modes. Note also that our proposed hidden variable distribution
function ${\cal P}(f_l, \tau_l ;f_r, \tau_r)$ reproduces the successful 
quantum--mechanical description of all the measurements in  Ref.~\cite{Go}.
More importantly, our {\it ad hoc} local realistic model also violates
the inequality measured there. This proves that the inequality tested
in Ref.~\cite{Go} is not a genuine Bell--inequality, which, by definition, has to be
satisfied in any local realistic approach. A similar criticism applies to the inequality derived
in Ref.~\cite{bf} for entangled $K^0\bar K^0$ pairs.
The failure of both discussions is due to the lack of an active intervention of the
experimenter.

%%%%%%%%%%%%%%%%%%%%%%%%%%%%%%%%%%%%%%%%%%%%%%%%%%%%%%%%%%%%%%%
\subsection{Proposals with active measurements in free space}
%%%%%%%%%%%%%%%%%%%%%%%%%%%%%%%%%%%%%%%%%%%%%%%%%%%%%%%%%%%%%%
The analogy between strangeness and linear polarization measurements has been 
exploited by many authors. In the analysis by Ghirardi {\it et al.}~\cite{ghi} 
one considers the $K^0\bar K^0$ state (\ref{statet}) and 
performs active joint strangeness measurements at two
different times on the left beam ($\tau_1$ and $\tau_2$) and at other two 
different times on the right beam ($\tau_3$ and $\tau_4$). The detection times
should be chosen at will and in accordance with the locality requirement.
The proposed inequality is again in the CHSH form \cite{chsh}:
\begin{equation}
\label{chsh-ghi}
\left|E_{\rm LR}(\tau_1,\tau_3)-E_{\rm LR}(\tau_1,\tau_4) +
E_{\rm LR}(\tau_2,\tau_3)+E_{\rm LR}(\tau_2,\tau_4)\right|\leq 2 ,
\end{equation}
where $E(\tau_r,\tau_r)$ is a correlation function which takes the value $+1$ when either
two $\bar K^0$'s or no $\bar K^0$'s are found in the left ($\tau_l$) and right ($\tau_r$)
measurements, and $-1$ otherwise:
\begin{equation}
E(\tau_l,\tau_r)\equiv P(Y,\tau_l;Y,\tau_r)+P(N,\tau_l;N,\tau_r)
-P(Y,\tau_l;N,\tau_r)-P(N,\tau_l;Y,\tau_r) .
\end{equation}
The probabilities entering this correlation function, where $Y$ (Yes) and $N$ (No) 
answer to the question whether a $K^0$ is detected at the considered time, can be 
obtained in QM from Eqs.~(\ref{4P1}) and (\ref{4P2}), and 
$E_{\rm QM}(\tau_l,\tau_r)=-\exp{\{-(\Gamma_L+\Gamma_S)(\tau_l+\tau_r)/2\}}
\cosÃ[\Delta m\, (\tau_l-\tau_r)]$.

Because of strangeness oscillations in free space along both kaon paths, choosing
among four different times corresponds to four different choices of measurement
directions in the photon case. In this sense, there is a total analogy and CHSH inequality
(\ref{chsh-ghi}) is a strict consequence of LR. Unfortunately,
this inequality is never violated by QM because strangeness
oscillations proceed too slowly and cannot compete with the more
rapid kaon weak decays. The conclusion is the same for the CHSH inequalities
derivable for the $B^0$--$\bar B^0$ and $D^0$--$\bar D^0$ meson systems \cite{hies-th,antiGo}.
On the contrary, genuine CHSH inequalities
violated by QM could be derived for $B^0_s\bar B^0_s$ pairs
if active flavour measurements were possible for these mesons.
As discussed in Refs.~\cite{gigo,dg}, Bell's inequalities exploiting
strangeness measurements at four different times can be violated by QM only if
a normalization of the observables to undecayed kaon pairs is employed. 
Unfortunately, the Bell's inequalities obtained with
such a normalization procedure are non--genuine \cite{antiGo}.

In Ref.~\cite{uchiyama}, Uchiyama derived the following Wigner--like 
inequality \cite{wig}:
\begin{equation}
\label{uchi}
P_{\rm LR}(K_S,K^0)\leq P_{\rm LR}(K_S,K_1) + P_{\rm LR}(K_1,K^0) ,
\end{equation}
for the entangled kaon state of Eq.~(\ref{entangled}).
The joint probabilities are assumed to be measured at a proper time $\tau=\tau_l=\tau_r$
very close to the instant of the pair creation, $\tau \to 0$; therefore the inequality
would eventually test noncontextuality rather than locality. Inserting the quantum--mechanical
probabilities  into Eq.~(\ref{uchi}), one obtains ${\cal R}e\, \epsilon \leq
|\epsilon|^2$, which is violated by the presently accepted value of $\epsilon$. 
Note that the proposed inequality involves passive measurements along a new, third
basis consisting of the two $CP$ eigenstates ($K_1$ and $K_2$). But the smallness of
$|\epsilon |$ and Eqs.~(\ref{cp}) preclude any realistic attempt 
of discriminating between lifetime ($K_S$ vs
$K_L$) and $CP$ ($K_1$ vs $K_2$) eigenstates.
In this sense, the interest of inequality (\ref{uchi}) reduces to that of a clear and
well defined  gedanken experiment.  

%%%%%%%%%%%%%%%%%%%%%%%%%%%%%%%%%%%%%%%%%%%%%%%%%%%%%%%%%%%%%%%%%%%%
\subsection{Proposals with active measurements and regenerators}
%%%%%%%%%%%%%%%%%%%%%%%%%%%%%%%%%%%%%%%%%%%%%%%%%%%%%%%%%%%%%%%%%%%%

The authors of Refs.~\cite{bn,abn}, while insisting on the
convenience of performing only unambiguous strangeness measurements, have substituted
the use of different times (as in Ref.~\cite{ghi})  
by the possibility of choosing among different kaon regenerators
to be inserted along the kaon path(s).
The well known regeneration effect can be interpreted as producing adjustable
``rotations''  in the kaon ``quasi--spin'' space analogous to the strangeness oscillations 
(i.e., quasi--spin oscillations in vacuum) in Ref.~\cite{ghi}, without
requiring additional time intervals. One can thus derive genuine Bell's inequalities,
violated by QM, for simultaneous left--right strangeness measurements.
The drawback of these analyses is that, up to now, they only refer
to thin regenerators and the predicted violations of
Bell's inequalities (below a few percent) are hardly observable.

Eberhard \cite{eberhard} considered the alternative option, 
based on $K_S$ vs $K_L$ identification,
for establishing a genuine Bell's inequality. He combined such measurements
in four experimental set--ups. In a first set-up,
the state (\ref{time}) is allowed to propagate in free space; its normalization is lost
because of weak decays, but its perfect antisymmetry is maintained. In the other three
set-ups, thick regenerators are asymmetrically located along one beam, or along the
other, or along both. An interesting inequality relating the number of $K_L$'s detected 
downstream from the production vertex and in
each experimental set-up is then derived from LR. It turns out to be
significantly violated by quantum--mechanical predictions. Unfortunately, these successful predictions
have some practical limitations, as already discussed by the author \cite{eberhard}. In
particular, they are valid for asymmetric $\phi$--factories (where the two neutral kaon
beams form a small angle), whose construction is not foreseen.

New forms of Bell's inequalities for
neutral kaons not affected by the drawbacks we have just mentioned have been 
derived in Ref.~\cite{BG}. Here, two kinds of active measurements, 
$K^0$ vs $\bar{K^0}$  and $K_S$ vs $K_L$, have been considered
in various alternative experimental set-ups with a thin regenerator fixed on the right beam 
as close as possible to the kaon--pair creation point. The proper time $\Delta \tau_r$
required by the  neutral kaon to cross the regenerator is assumed to be short
enough ($\Delta \tau_r \ll \tau_S$) to neglect weak decays. Then free space propagation
is allowed up to a proper time $T$, with $\tau_S \ll T \ll \tau_L$. The normalization
to surviving pairs leads then to the non--maximally entangled state:
\begin{equation}
\label{stateN}
|\Phi \rangle= {1 \over \sqrt{2 + |R|^2}} \left[ |K_S\rangle_l |K_L\rangle_r  
- |K_L\rangle_l |K_S\rangle_r + R |K_L\rangle_l |K_L\rangle_r \right],
\end{equation}
where
\begin{equation}
%\label{R}
R \equiv  -r e^{-i\left(\Delta m -{i \over 2}\Delta \Gamma\right)T} 
\end{equation}
and
\begin{equation}
\label{r}
r \equiv i{\pi \nu \over m_K}(f - \bar{f}) \Delta \tau_r =
i{\pi \nu  \over p_K}(f - \bar{f}) d  
\end{equation}
is the regeneration parameter. In Eq.~(\ref{r}),  
$m_K$ is the average neutral kaon mass, $p_K$ the kaon momentum,
$f$ ($\bar{f}$) the ${K^0}$--nucleus ($\bar{K^0}$--nucleus) forward
scattering amplitude, $\nu$ the density of scattering centers of the homogeneous
regenerator whose total thickness is $d$. The state (\ref{stateN}) describes all kaon
pairs with both left and right partners surviving up to a common proper time $T$. 

At this point, alternative measurements of strangeness or lifetime
will be performed on each one of these kaon pairs (\ref{stateN}) according to the strategies
for active measurement procedures illustrated in Section \ref{nms}. 
Care has to be taken to choose
$T$ large enough to guarantee the space--like separation between left and right
measurements. Locality excludes then any influence from the experimental set-up
encountered by one member of the kaon pair at time $T$ on the behaviour of
its other--side partner between $T$ and $T+ \Delta \tau$. 
For kaon pairs from $\phi$ decays, moving at $\beta \simeq 0.22$, and using an interval
time $\Delta \tau = 4.8\, \tau_S$ for the lifetime identification, 
this implies $T > (\beta^{-1}-1) \Delta \tau /2= 8.7\, 
\tau_S$, with a considerable reduction of the total kaon sample; a reduction which is
much more moderate for more relativistic kaons as in $p \bar{p}$ annihilations. 

The requirements (1)--(4) of Section \ref{req}
for deriving genuine Bell's inequalities are thus fulfilled and
one can write several inequalities. Among these, we first discussed 
\cite{BG} an homogeneous Clauser and Horne (CH) inequality \cite{clh} 
which was substantially violated by QM. 
Note moreover that, as discussed in Ref.~\cite{clh}, homogeneous CH
inequalities have the advantage of being independent of the normalization of the total sample
of pairs involved and are thus easier to test than non--homogeneous ones. 
More recently, in Ref.~\cite{BG1} we have improved the analysis of Ref.~\cite{BG} by 
applying Hardy's proof without inequalities of Bell's 
theorem \cite{Ha93} to the state (\ref{stateN}). 

Let us concentrate on the proof of Ref.~\cite{BG1}. Neglecting $CP$--violation and
$K_L$--$K_S$  misidentification effects, from state (\ref{stateN}) with $R=-1$ 
(called Hardy's state) one obtains the following quantum--mechanical predictions:
\begin{eqnarray}
\label{non-zero}
P_{\rm QM}(K^0,\bar{K}^0)&=&\frac{\eta\, \bar \eta}{12} , \\
\label{zero1}
P_{\rm QM}(K^0,K_L)&=&0 , \\
\label{zero2}
P_{\rm QM}(K_L,\bar{K}^0)&=&0 , \\
\label{quasi-zero}
P_{\rm QM}(K_S,K_S)&=& 0 ,
\end{eqnarray}
where $\eta$ ($\bar \eta$) is the $K^0$ ($\bar K^0$) overall detection efficiency.
It is found that the necessity to reproduce, under LR, equalities 
(\ref{non-zero})--(\ref{zero2}) requires:
\begin{equation}
P_{\rm LR}(K_S,K_S) \geq P_{\rm LR}(K^0,\bar{K}^0)=\frac{\eta\, \bar \eta}{12},
\end{equation}
which contradicts Eq.~(\ref{quasi-zero}). In principle, this allows for an 
``all--or--nothing'' Hardy--like test of LR vs QM. In Ref.~\cite{BG1} it was concluded
that, by requiring a perfect discrimination between $K_S$ and $K_L$ states, an
experiment measuring the joint probabilities of
Eqs.~(\ref{non-zero})--(\ref{quasi-zero}) closes the efficiency loophole even for
infinitesimal values of the strangeness detection efficiencies $\eta$ and $\bar \eta$.
However, since $K_L$--$K_S$ misidentifications 
(due to the finite value of $\Gamma_S/\Gamma_L\simeq 579$) do not permit an ideal
lifetime measurement even when the detection efficiency $\eta_\tau$ for the kaon decay
products is $100$\% \cite{genovese}, the original proposal must be reanalyzed paying  
particular attention to the inefficiencies involved in the real test.

Retaining the effects due to the $K_S$--$K_L$ misidentification, 
from Eq.~(\ref{stateN}) with $R=-1$ one obtains (see the Appendix for details):
\begin{eqnarray}
\label{SS0}
P_{\rm QM}(K^0,\bar{K}^0) &=& \frac{\eta \bar \eta}{12} , \\
\label{SL0}
P_{\rm QM}(K^0,K_L) &=&  
6.77 \times 10^{-4} \eta \, \eta_{\tau} , \\
\label{LS0}
P_{\rm QM}(K_L,\bar K^0) &=& 
6.77 \times 10^{-4} \bar \eta \, \eta_{\tau} , \\
\label{LL0}
P_{\rm QM}(K_S,K_S)
&=& 1.19 \times 10^{-5} \eta_\tau^2, 
\end{eqnarray}
which replace the results of Eqs.~(\ref{non-zero})--(\ref{quasi-zero}) 
and where $\eta_\tau$ is the efficiency for the detection of the kaon decay products. 
In the standard Hardy--like proof of non--locality \cite{Ha93},
the probabilities corresponding to our (\ref{SL0}), (\ref{LS0}) and (\ref{LL0})
are perfectly vanishing. In our realistic case they are very small but not zero. 
Nevertheless, this does not prevent us from deriving a contradiction between LR and QM.
Indeed, as proved in Ref.~\cite{EbRo}, the well known criterion of physical reality of
Einstein, Podolsky and Rosen \cite{epr} can be
generalized to include predictions made with {\it almost} certainty,
as it is required in our case due to the nonvanishing values of
probabilities (\ref{SL0})--(\ref{LL0}).
The proof of non--locality without inequalities of Ref.~\cite{BG1}
remains unchanged, and one obtains again the condition
$P_{\rm LR}(K_S,K_S) \geq P_{\rm LR}(K^0,\bar{K}^0)$, which is incompatible
with QM if the detection efficiencies verify the inequality:
\begin{equation}
\eta\, \bar \eta > 1.4 \times 10^{-4}\eta^2_{\tau} .
\end{equation}

In order to prove whether LR is refuted by Nature, the quantities of 
Eqs.~(\ref{SS0})--(\ref{LL0}) 
must be measured. One thus has to confirm probabilities whose values, in QM,
are almost zero. The difficulties associated to ``almost null'' measurements can be
overcome if one employs an inequality \cite{hardy94} involving
all the probabilities needed in the proof of Bell's theorem
without inequalities. The use of an inequality also allows for small
deviations (existing in real experiments) around the value $R=-1$ required to
prepare our Hardy's state. 
What we need is the following Eberhard's inequality \cite{Eb}:
\begin{eqnarray}
\label{hardy}
\hskip -8mm
H_{\rm LR}&\equiv& \frac{P_{\rm LR}(K^0,\bar{K}^0)}{P_{\rm LR}(K^0,K_L)
+ P_{\rm LR}(K_S,K_S) + P_{\rm LR}(K_L,\bar{K}^0)+P(K^0,U_{\rm Lif})+P(U_{\rm Lif},\bar K^0)} \leq 1 .
\end{eqnarray}
Essentially, it is a different writing of the
following homogeneous CH inequality \cite{clh}:
\begin{equation}
\label{clauser-horne}
Q_{\rm LR}\equiv
\frac{P_{\rm LR}(K_S,\bar K^0)-P_{\rm LR}(K_S,K_S)+P_{\rm LR}(K^0,\bar K^0)
+P_{\rm LR}(K^0,K_S)}{P_{\rm LR}(K^0,*)+P_{\rm LR}(*,\bar K^0)} \leq 1 , \nonumber
\end{equation}
where
\begin{eqnarray}
\label{star}
P_{\rm LR}(K^0,*) &=& P_{\rm LR}(K^0,K_S) + P_{\rm LR}(K^0,K_L) + P_{\rm LR}(K^0,U_{\rm Lif}), \\
P_{\rm LR}(*,\bar K^0) &=& P_{\rm LR}(K_L,\bar K^0) + P_{\rm LR}(K_S,\bar K^0)+
P_{\rm LR}(U_{\rm Lif},\bar K^0) , \nonumber
\end{eqnarray}
and the argument $U_{\rm Lif}$ refers to failures in lifetime detection. 
Both inequalities are actually derivable from LR for any value of $R$. However,
Hardy's proof leads to inequality (\ref{hardy}) only for Hardy's state ($R=-1$).
Note that the probabilities containing lifetime undetection, 
whose expressions in QM are: 
\begin{eqnarray}
\label{extra1}
P_{\rm QM}(K^0,U_{\rm Lif})&=&\frac{1}{6} \eta \left(1-\eta_{\tau} \right),  \\
\label{extra2}
P_{\rm QM}(U_{\rm Lif},\bar K^0)&=&\frac{1}{6} \bar \eta \left(1-\eta_{\tau} \right) , 
\end{eqnarray}
appear in Eberhard's inequality (\ref{hardy}) and in the single--side probabilities of Eq.~(\ref{star}).
Note also that the previous Eberhard's and CH inequalities have been obtained {\it
without invoking supplementary assumptions} on undetected events. They are both genuine
Bell's inequalities and provide the same restrictions on the efficiencies $\eta$, $\bar
\eta$ and $\eta_{\tau}$ required for a detection loophole free experiment. 

In order to discuss the feasibility of such an experiment let us start considering a
few ideal cases. Assume first that  perfect discrimination
between $K_S$ and $K_L$ were always possible ($\eta_\tau=1$ and $p_L=p_S=1$, see Appendix);  
one could then make a conclusive test of LR for any nonvanishing
values of $\eta$ and $\bar \eta$: 
$H^{\eta_\tau= p_L=p_S=1}_{\rm QM}\to \infty$, 
$\forall\, \eta, \bar \eta\neq 0$. In a second ideal case with no undetected
events, i.e. with $\eta=\bar \eta=\eta_{\tau}=1$, the inequalities are strongly violated
by QM: $H^{\eta= \bar \eta = \eta_{\tau}=1}_{\rm QM}\simeq 60.0$, 
$Q^{\eta= \bar \eta = \eta_{\tau}=1}_{\rm QM}\simeq 1.25$, 
even if one allows for unavoidable $K_S$ and $K_L$ misidentifications. Finally, 
assuming that only the detection efficiency of kaon decay products is ideal
($\eta_{\tau}=1$), for $\eta=\bar \eta$ ($\eta=\bar \eta/2$), Eberhard's and CH
inequalities are contradicted by QM whenever $\eta> 0.023$ ($\eta> 0.017$).

\begin{figure}[t]
\begin{center}
\mbox{\epsfig{file=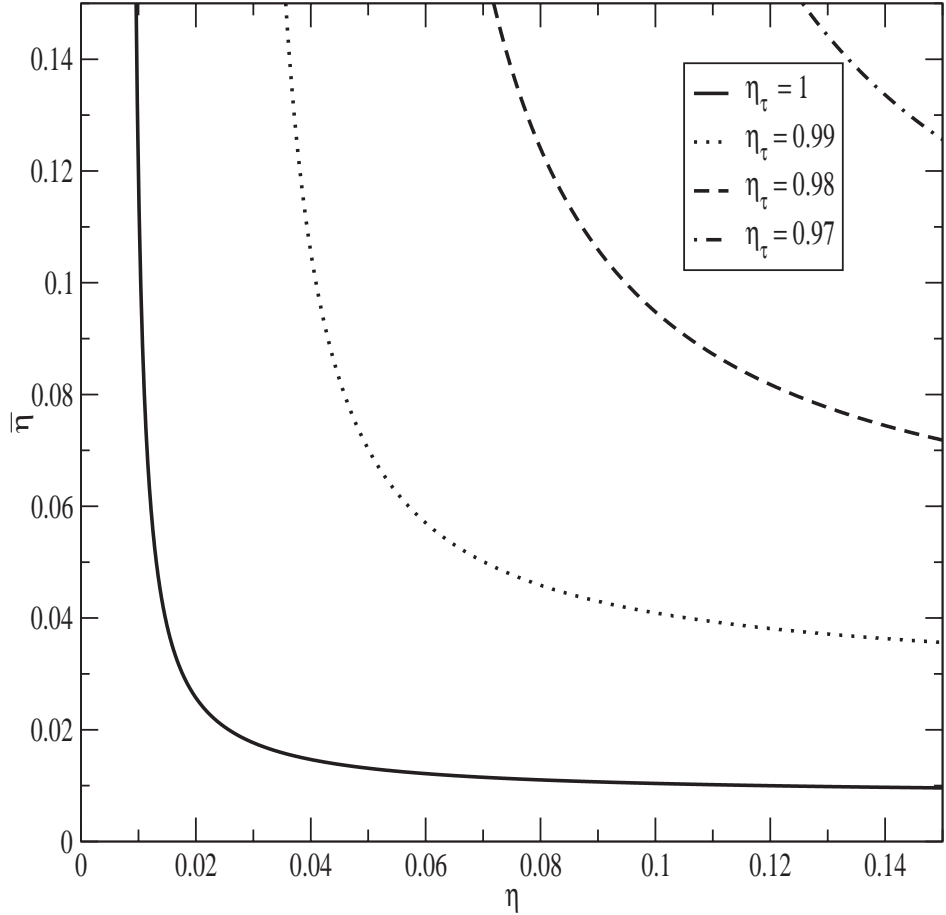,width=.75\textwidth}}
\vskip 2mm
\caption{The four curves (corresponding to
$\eta_{\tau}=1,0.99,0.98$ and $0.97$)
provide the values of $\eta$ and $\bar \eta$ for which
$H_{\rm QM}=Q_{\rm QM}=1$ using Hardy's state.
QM violates inequalities (\ref{hardy}) and (\ref{clauser-horne}) for
values of $\eta$ and $\bar \eta$ situated above the corresponding curve.}
\label{figura}
\end{center}
\end{figure}

Let us now consider more realistic situations with 
small and achievable values of $\eta$ and $\bar \eta$. This implies that we
have to consider large, but still realistic, decay--product
detection efficiencies such as $\eta_{\tau}=0.97$, $0.98$, $0.99$ and, ideally, $1$.
For each $\eta_{\tau}$, the values of $\eta$ and $\bar \eta$ that permit a detection
loophole free test ($H_{\rm QM}$, $Q_{\rm QM}>1$) lie above the corresponding curve
plotted in Fig.~\ref{figura}.
As expected, when $\eta_{\tau}$ decreases, the region
of $\eta$ and $\bar\eta$ values which permits a conclusive test
diminishes and larger values of $\eta$ and $\bar\eta$ are required.
Note, however, that the strangeness detection efficiencies required for a conclusive
test of  LR vs QM with neutral kaons are considerably smaller than the limit
($\eta_0=0.67$) deduced  by Eberhard \cite{Eb} for non--maximally entangled photon
states. The  values for $\eta$ and $\bar \eta$ required by the test we have proposed
seem to be not far from the present experimental capabilities.

%%%%%%%%%%%%%%%%%%%%%%%%%%%%%%%%%%%%%%%%%%%%%%
\section{Conclusions}
%%%%%%%%%%%%%%%%%%%%%%%%%%%%%%%%%%%%%%%%%%%%%%

A series of recent proposals aiming to perform 
Bell's inequality tests with entangled pairs of pseudoscalar mesons have been
discussed. This includes, in particular, pairs of neutral kaons or $B$--mesons. The
relativistic velocities of these mesons and their strong interactions seem to offer the
possibility of simultaneously closing the so--called locality and detection loopholes.
The real situation, however, is not a simple one. 

In several proposals, the measurements required to perform a Bell--test consist in
identifying the flavour of each meson via its observed decay mode. The 
inequalities so derived are not a consequence of LR and, in this sense, cannot provide
Bell--tests of LR vs QM. The reason is that the observed meson decays correspond to passive
flavour measurements ---with no choice for the experimenter--- in such a way that a local
realistic model can always be constructed reproducing all the probabilities predicted by QM.

Other proposals suffer from the difficulties coming from the fact that the
number of different complementary measurements on pseudoscalar mesons is very
small. For neutral kaons, for instance, they essentially reduce to
strangeness and lifetime measurements. A situation which can be improved if the well
known effects of kaon regeneration are taken into account. 

Indeed, a series of papers have proposed Bell--tests with neutral kaons
using kaon regeneration.
On the one hand, this amounts to an effective increase of the number of
mutually exclusive measurements one can perform. On the other, by changing or removing
the regenerators the active presence of the experimenter is guaranteed. A final
difficulty could still remain: the low efficiency of some of these neutral kaon
measurements. A detailed analysis suggests that  
a detection loophole free Bell--test with neutral kaons would require a few \% 
strangeness detection efficiencies and very high efficiencies for the detection
of the kaon decay products. Both requirements seem achievable with present day
technology. 

%%%%%%%%%%%%%%%%%%%%%%%%%%
\section*{Appendix}
%%%%%%%%%%%%%%%%%%%%%%%%
In Ref.~\cite{BG1}, $K_S$'s states at time $T$ are identified through decay events 
taking place between times $T$ and $T+\Delta \tau$; similarly, $K_L$'s states are
identified as kaons decaying after time $T+\Delta \tau$. For $\Delta \tau = 4.8\, \tau_S$,
the probabilities for correct $K_S$ and $K_L$ identifications are:
\begin{equation}
\label{8}
p_S\equiv 1-\exp{(-4.8)}=p_L\equiv \exp{(-4.8/579)}=0.9918 .
\end{equation}
and misidentifications are thus at the level of some $8$ per thousand.
 
One can further reduce these misidentifications by considering
not only the kaon decay time but also the decay channel. Neglecting $K_S$ and $K_L$
branching ratios smaller that $10^{-5}$, decays into $\pi\pi\pi$ identify $K_L$'s
and only semileptonic and $\pi\pi$ channels are accessible to both $K_S$
and $K_L$ \cite{PDG}:
$BR(K_L\to \pi e \nu_e\,\, {\rm or}\,\, \pi \mu \nu_\mu)=0.6600$,
$BR(K_L\to \pi\pi)=0.0030$, 
$BR(K_S\to \pi e \nu_e\,\, {\rm or}\,\, \pi \mu \nu_\mu)=0.0011$
and $BR(K_S\to \pi\pi)=0.9989$.
However, semileptonic decays have to be assigned to $K_L$'s decays for any
decay time (this introduce a misidentification, equal to
$BR(K_S\to \pi e \nu_e\,\, {\rm or}\,\, \pi \mu \nu_\mu)=1.1\times 10^{-3}$, in the $K_S$ identification).
Indeed, the probability that a $K_L$ decays semileptonically in a time interval $\Delta
\tau$ after $T$ is larger than the probability corresponding to a $K_S$, for any value
of $\Delta \tau$.  A decay into $\pi\pi$ occurring between $T$ and $T+5.82\, \tau_S$
(after $T+5.82\, \tau_S$) has to be assigned to a $K_S$ ($K_L$). In fact,
the probability that a $K_S$ [$K_L$], which is alive at time $T$, decays into $\pi\pi$
after $T+\Delta \tau$ is $P_S(\Delta \tau)=\exp(-\Delta \tau/\tau_S)BR(K_S\to \pi\pi)$
[$P_L(\Delta \tau)=\exp(-\Delta \tau/\tau_L)BR(K_L\to \pi\pi)$] and
$P_L(\Delta \tau)$
is larger (smaller) than $P_S(\Delta \tau)$ for $\Delta \tau > 5.82\, \tau_S$
($\Delta \tau < 5.82\, \tau_S$). The probabilities that $K_S$'s and $K_L$'s are actually
identified as $K_S$'s and $K_L$'s are thus:
\begin{eqnarray}
\label{mejor}
p_S&=&1-BR(K_S\to \pi e \nu_e\,\, {\rm or}\,\, \pi \mu \nu_\mu)
-BR(K_S\to \pi\pi)\exp(-5.82) \\
&=&BR(K_S\to \pi\pi)[1-\exp(-5.82)]=0.99594 , \nonumber \\
p_L&=&1-BR(K_L\to \pi\pi)[1-\exp(-5.82/579)]=0.99997 \nonumber ,
\end{eqnarray}
thus improving the lifetime identification with respect to the method of Eq.~(\ref{8}).

Retaining the effects due to the $K_S$--$K_L$ misidentification
($CP$--violation and the nonorthogonality of $|K_L\rangle$ and $|K_S\rangle$ 
can indeed be neglected),
from Eq.~(\ref{stateN}) with $R=-1$ we obtain:
\begin{eqnarray}
\label{SS0A}
P_{\rm QM}(K^0,\bar{K}^0) &=& \frac{\eta \bar \eta}{12} , \\
\label{SL0A}
P_{\rm QM}(K^0,K_L) &=& |\langle K^0 K_S|\Phi \rangle|^2 \eta \, \eta_{\tau} (1- p_S)=
\frac{1}{6}\eta \, \eta_{\tau} (1- p_S) , \\
\label{LS0A}
P_{\rm QM}(K_L,\bar K^0) &=& |\langle K_S \bar K^0|\Phi \rangle|^2 \bar \eta \, \eta_{\tau} (1-p_S)=
\frac{1}{6}\bar \eta \, \eta_{\tau} (1- p_S) , \\
\label{LL0A}
P_{\rm QM}(K_S,K_S)
&=& \frac{2}{3}\eta_\tau^2\left\{p_S(1-p_L) -BR(K_S\to \pi \pi)\, BR(K_L\to \pi \pi)\, 
\frac{\Gamma_S\, \Gamma_L} {\Gamma^2+\Delta m^2} \right. \\
&& \left. \times \left[1-2e^{-5.82\frac{\Gamma}{\Gamma_S}} 
\cos \left(5.82 \frac{\Delta m}{\Gamma_S}\right) 
+e^{-2\times 5.82 \frac{\Gamma}{\Gamma_S}}\right]\right\}, \nonumber
\end{eqnarray}
from which the numerical values of Eqs.~(\ref{SL0})--(\ref{LL0}) follow via 
Eq.~(\ref{mejor}) and Ref.~\cite{PDG}. 

In Eq.~(\ref{SL0A}) [(\ref{LS0A})] semileptonic $K_S$ decay events on 
the right (left) and $K_S$ states surviving up to $T+5.82\, \tau_S$
are wrongly assumed as coming from $K_L$'s. The derivation of
Eq.~(\ref{LL0A}) deserves some comment. Since $K_S$'s are identified through their
$\pi \pi$ decays occurring between times $T$ and $T+5.82\, \tau_S$,
experimentally one has to measure the following double differential rate:
\begin{equation}
\Gamma(\pi \pi, \tau_l; \pi \pi, \tau_r)=\int d\Omega_l \int d\Omega_r 
\left|A(\pi \pi, \tau_l; \pi \pi, \tau_r)\right|^2 ,
\end{equation}
where the integrations are over the phase space for the decay product states
and $0\leq \tau_l, \tau_r\leq 5.82\, \tau_S$. The corresponding amplitude is obtained from 
Eq.~(\ref{stateN}) with $R=-1$ as:
\begin{equation}
A(\pi \pi, \tau_l; \pi \pi, \tau_r)=\frac{1}{\sqrt{3}}
\langle \pi \pi|T|K_S \rangle \langle \pi \pi|T|K_L \rangle
\left[e^{-i\lambda_S \tau_l-i\lambda_L \tau_r}-e^{-i\lambda_L \tau_l-i\lambda_S \tau_r}\right] ,
\end{equation}
where we have neglected the small contribution coming from the $|K_L\rangle_l |K_L\rangle_r$
part of the state and $\lambda_{L,S}=m_{L,S}-i\,\Gamma_{L,S}/2$. The joint probability 
(\ref{LL0A}) is then computed with the following relation:
\begin{equation}
P_{\rm QM}(K_S,K_S)=\int_0^{5.82\, \tau_S} d\tau_l \int_0^{5.82\, \tau_S} d \tau_r \,
\Gamma(\pi \pi, \tau_l;\pi \pi, \tau_r) .
\end{equation}

%%%%%%%%%%%%%%%%%%%%%%%%%%%
\section*{Acknowledgements}
%%%%%%%%%%%%%%%%%%%%%%%%%%%
Work partly supported by EURIDICE HPRN--CT--2002--00311, MIUR 2001024324\_007, INFN
and DGICYT BFM-2002-02588.
This work is also partly supported by the Ramon y Cajal program (R.E.).
%%%%%%%%%%%%%%%%%%%%%%%%%%%%%%%


\begin{thebibliography}{100}
%%%%%%%%%%%%%%%%%%%%%%%%%%%%%%%

\bibitem{epr}
A. Einstein, B. Podolsky and N. Rosen, {\em Phys. Rev.} {\bf 47}, 777 (1935).

\bibitem{Bo35}
N. Bohr, {\em Phys. Rev.} {\bf 48}, 696 (1935).

\bibitem{bell}
J. Bell, {\em Physics} {\bf 1}, 195 (1964).

\bibitem{Redhead}
M.~Redhead, {\em Incompleteness, non--locality and realism}
(Oxford University Press, Oxford, 1990).

\bibitem{wig}
E. P. Wigner, {\em Am. J. Phys.} {\bf 38}, 1005 (1970).

\bibitem{Bell2}
J.S.~Bell, {\em Speakable and unspeakable in quantum mechanics
(collected papers on quantum philosophy)}, (Cambridge University Press, 1987).

\bibitem{chsh}
J. F. Clauser, M. A. Horne, A. Shimony and R. A. Holt, {\em Phys. Rev. Lett.}
{\bf 23}, 880 (1969).

\bibitem{clh}
J. F. Clauser and M. A. Horne, {\em Phys. Rev.} {\bf D 10}, 526 (1974).

\bibitem{As82}
%A. Aspect, P. Grangier and G. Roger, {\em Phys. Rev. Lett.} {\bf 47}, 460 (1981);
%ibid. {\bf 49}, 91 (1982);
A. Aspect, J. Dalibard and G. Roger, {\em Phys. Rev. Lett.} {\bf 49}, 1804 (1982).

\bibitem{review1}
J. F. Clauser and A. Shimony, {\em Rep. Prog. Phys.} {\bf 41}, 1881 (1978).

\bibitem{We98}
G. Weihs, T. Jennewein, C. Simon, H. Weinfurter and A. Zeilinger,
{\em Phys. Rev. Lett.} {\bf 81}, 5039 (1998).

\bibitem{Ti99}
W. Tittel, J. Brendel, N. Gisin and H. Zbinden, {\em Phys. Rev.} {\bf A 59},
4150 (1999); W. Tittel, J. Brendel, H. Zbinden and N. Gisin, 
{\em Phys. Rev. Lett.} {\bf 81}, 3563 (1998).

\bibitem{BZ}
R.~A.~Bertlmann and A.~Zeilinger (Eds.),
{\em Quantum (Un)speakables -- From Bell to Quantum Information},
(Springer, Berlin, 2002).

\bibitem{Ro01}
M.~A.~Rowe, D.~Kielpinski, V.~Meyer, C.~A.~Sackett, W.~M.~Itano,
C.~Monroe, and D.~J.~Wineland, {\em Nature} {\bf 409}, 791 (2001).

\bibitem{vaidman}
L. Vaidman, {\em Phys. Lett.} {\bf A 286}, 241 (2001).

\bibitem{santos}
E. Santos, {\em Phys. Lett.} {\bf A 327}, 33 (2004); quant-ph/0410193; quant-ph/0103062.

\bibitem{pearle}
P. Pearle, {\em Rep. Rev.}  {\bf D 2}, 1418 (1970).

\bibitem{GM}
A. Garg and N. D. Mermin, {\em Phys. Rev.} {\bf D 35}, 3831 (1987).

\bibitem{Eb}
P. H. Eberhard, {\em Phys. Rev.} {\bf A 47}, R747 (1993).

\bibitem{santos-models}
E. Santos, {\em Phys. Rev.} {\bf A 46}, 3646 (1992); {\em Phys. Lett.}
{\bf A 212}, 10 (1996); N. Gisin and B. Gisin, {\em Phys. Lett.} {\bf A 260}, 323 (1999).

\bibitem{ghi}
G. C. Ghirardi, R. Grassi and T. Webern, in {\em Proceedings of the
Workshop on Physics and Detectors for Da$\Phi$ne},
edited by G. Pancheri (INFN, Laboratori Nazionali di Frascati,
Frascati, Italy, 1991) p. 261.
%G. C. Ghirardi, R. Grassi and R. Regazzon, in {\em The Da$\Phi$ne Handbook},
%edited by L. Maiani, G. Pancheri and N. Paver (INFN, Laboratori
%Nazionali di Frascati, Frascati, Italy, 1992) p. 283.

\bibitem{eberhard}
P. H. Eberhard, {\em Nucl. Phys.} {\bf B 398}, 155 (1993).

\bibitem{domenico}
A. Di Domenico, {\em Nucl. Phys.} {\bf B 450}, 293 (1995).

\bibitem{uchiyama}
F. Uchiyama, {\em Phys. Lett.} {\bf A 231}, 295 (1997).

\bibitem{selleri}
F. Selleri, {\em Phys. Rev.} {\bf A 56}, 3493 (1997); 
R. Foadi and F. Selleri, {\em Phys. Lett.} {\bf B 461}, 123 (1999);
{\em Phys. Rev.} {\bf A 61}, 012106 (2000).

\bibitem{AS}
A. Afriat and F. Selleri, {\em The Einstein, Podolsky and Rosen paradox
in atomic, nuclear and particle physics}
(Plenum Press, New York, 1998).

\bibitem{bf}
F. Benatti and R. Floreanini, {\em Phys. Rev.} {\bf D 57}, R1332 (1998); 
{\em Eur. Phys. J.} {\bf C 13}, 267 (2000).

\bibitem{berthies}
R. A. Bertlmann, W. Grimus and B. C. Hiesmayr, {\em Phys. Rev.} {\bf D 60}, 114032 (1999); 
{\em Phys. Lett.} {\bf A 289}, 21 (2001);
B. C. Hiesmayr, {\em Found. Phys. Lett.} {\bf 14}, 231 (2001); 
R. A. Bertlmann and B. C. Hiesmayr, {\em Phys. Rev.} {\bf A 63}, 062112 (2001); 
R. A. Bertlmann, K. Durstberger and B. C. Hiesmayr, {\em Phys. Rev.} {\bf A 68}, 012111 (2003).

\bibitem{bn}
A. Bramon and M. Nowakowski, {\em Phys. Rev. Lett.} {\bf 83}, 1 (1999).

\bibitem{abn}
B. Ancochea, A. Bramon and M. Nowakowski, {\em Phys. Rev.} {\bf D 60},
094008 (1999).

\bibitem{gigo}
N. Gisin and A. Go, {\em Am. J. Phys.} {\bf 69}, 264 (2001).

\bibitem{dg}
R. Dalitz and G. Garbarino,  {\em Nucl. Phys.} {\bf B 606}, 483 (2001).
                                                                                                              
\bibitem{gnp}
M. Genovese, C. Novero and E. Predazzi, {\em Phys. Lett.} {\bf B 513}, 401 (2001); 
{\em Found. Phys.} {\bf 32}, 589 (2002).
   
\bibitem{BG}
A. Bramon and G. Garbarino, {\em Phys. Rev. Lett.} {\bf 88}, 040403 (2002). 
                                                    
\bibitem{BG1}
A.~Bramon and G.~Garbarino, {\em Phys. Rev. Lett.} {\bf 89}, 160401 (2002).

\bibitem{hies-th}
B. C. Hiesmayr, Ph.D. Thesis, University of Vienna, 2002.

\bibitem{genovese}
M. Genovese, {\em Phys. Rev.} {\bf A 69}, 022103 (2004).

\bibitem{Bert}
R. A. Bertlmann, quant-ph/0410028.

\bibitem{DH}
A. Datta and D. Home, {\em Phys. Lett.} {\bf A 119}, 3 (1986).

\bibitem{Se00}
A. Pompili and F. Selleri, {\em Eur. Phys. J.} {\bf C 14}, 469 (2000).

\bibitem{Go}
A.~Go, {\em J. Mod. Optics} {\bf 51}, 991 (2004); quant-ph/0310192.

\bibitem{antiGo}
R. A. Bertlmann, A. Bramon, G. Garbarino and B. C. Hiesmayr,
{\em Phys. Lett.} {\bf A 332}, 355 (2004).

\bibitem{antiGo2}
A. Bramon, R. Escribano and G. Garbarino, quant-ph/0410122.

\bibitem{daphne}
{\em The Second Da${\Phi}$ne Physics Handbook}
edited by L. Maiani, G. Pancheri and N. Paver (INFN, Laboratori
Nazionali di Frascati, Frascati, Italy, 1995).

\bibitem{belle}
S. Kurokawa and E. Kikutani, {\em Nucl. Instrum. Meth.} {\bf A 499}, 1 (2003).
%Belle Collaboration, {\em Technical Design Report}, KEK--R--95--1 (1995).

\bibitem{CPLEAR}
A. Apostolakis et al., {\em Phys. Lett.} {\bf B 422}, 339 (1998).

\bibitem{PDG}
S. Eidelman et al. (Particle Data Group), {\em Phys. Lett.} {\bf B 592}, 1 (2004).
%K.~Hagiwara et al., {\em Phys. Rev.} {\bf D 66}, 010001 (2002).

\bibitem{kabir}
P. K. Kabir, {\em The CP Puzzle} (Academic Press, London, 1968).

\bibitem{BGHPR}
A.~Bramon, G.~Garbarino and B.~C.~Hiesmayr, {\em Phys. Rev.}
{\bf A 69}, 062111 (2004).

\bibitem{CPLEARreview}
A.~Angelopoulos et al., {\em Phys. Rept.} {\bf 374}, 165 (2003);
{\em Phys. Lett.} {\bf 503}, 49 (2001): {\bf 444}, 38 (1998).

\bibitem{BGH2}
A.~Bramon, G.~Garbarino and B.~C.~Hiesmayr, {\em Eur. Phys. J.}
{\bf C 32}, 377  (2004).

\bibitem{BGH}
A.~Bramon, G.~Garbarino and B.~C.~Hiesmayr, {\em Phys. Rev. Lett.}
{\bf 92}, 020405 (2004).

\bibitem{kloe-int}
A. Di Domenico, hep-ex/0312032, published in eConf C0309101:THWP007,2003.

\bibitem{Kasday}
L.~Kasday, in {\em Foundations of Quantum Mechanics}, B. d'Espagnat ed.
(New York, Academic Press, 1971), p.195. Proceedings of the
International School of Physics `Enrico Fermi', Course IL.

\bibitem{Ha93}
L. Hardy, {\em Phys. Rev. Lett.} {\bf 68}, 2981 (1992);
{\em Phys. Rev. Lett.} {\bf 71}, 1665 (1993).

\bibitem{EbRo}
P. H. Eberhard and P. Rosselet, Universite de Lausanne Report No. IPNL--93--3, 1993;
{\em Found. Phys.} {\bf 25}, 91 (1995).

\bibitem{hardy94}
L. Hardy, {\em Phys. Rev. Lett.} {\bf 73}, 2279 (1994);
N. D. Mermin, {\em Am. J. Phys.} {\bf 62}, 880 (1994);
A. Garuccio, {\em Phys. Rev.} {\bf A 52}, 2535 (1995).

\end{thebibliography}
\end{document}